\documentclass[manuscript,screen]{acmart}
\AtBeginDocument{%
  }

\setcopyright{acmlicensed}
\copyrightyear{2018}
\acmYear{2018}
\acmDOI{XXXXXXX.XXXXXXX}
\acmConference[Conference acronym 'XX]{Make sure to enter the correct
  conference title from your rights confirmation email}{June 03--05,
  2018}{Woodstock, NY}
\acmISBN{978-1-4503-XXXX-X/2018/06}
\usepackage{booktabs}

\usepackage{amssymb}
\usepackage{tabularx}
\usepackage{pifont}
\usepackage[most]{tcolorbox}
\usepackage{xcolor}
\usepackage{tikz}
\usepackage{xspace}
\definecolor{gree}{HTML}{1e8449}
\newcommand{\fish}[1]{\textcolor{gree}{#1}\xspace}

\newcommand{\RomanNum}[1]{\MakeUppercase{\romannumeral #1}}
\newcommand{\etal}{\textit{et al.}\xspace}

\definecolor{HMColor}{RGB}{128, 164, 146}
\definecolor{HSColor}{RGB}{114, 142, 253}
\newcommand*\HM[1]{\tikz[baseline=(char.base)]{
            \node[shape=rectangle,fill=HMColor, text=white, inner sep= 1pt,minimum size=8pt,rounded corners=1pt] (char) {\textbf{#1}}}}
\newcommand*\HS[1]{\tikz[baseline=(char.base)]{
            \node[shape=rectangle,fill=HSColor, text=white, inner sep= 1pt,minimum size=8pt,rounded corners=1pt] (char) {\textbf{#1}}}}
\begin{document}

\title{Human in the Loop for Fuzz Testing: Literature Review and the Road Ahead}

\author{Jiongchi Yu}
\affiliation{%
  \institution{Singapore Management University}
  \state{Singapore}
  \country{Singapore}
}
\authornote{Both authors contributed equally to this research.}
\email{jcyu.2022@phdcs.smu.edu.sg}
\author{Xiaolin Wen}
\authornotemark[1]
\email{xiaolin004@e.ntu.edu.sg}
\affiliation{%
  \institution{Nanyang Technological University}
  \state{Singapore}
  \country{Singapore}
}

\author{Sizhe Cheng}
\email{@e.ntu.edu.sg}
\affiliation{%
  \institution{Nanyang Technological University}
  \state{Singapore}
  \country{Singapore}
}

\author{Xiaofei Xie}
\email{xfxie@smu.edu.sg}
\affiliation{%
  \institution{Singapore Management University}
  \state{Singapore}
  \country{Singapore}
}

\author{Qiang Hu}
\authornote{Corresponding authors}
\email{qianghu@tju.edu.cn}
\affiliation{%
  \institution{Tianjin University}
  \state{Tianjin}
  \country{China}
}



\author{Yong Wang}
\authornotemark[2]
\email{yong-wang@ntu.edu.sg}
\affiliation{%
  \institution{Nanyang Technological University}
  \state{Singapore}
  \country{Singapore}
}


\renewcommand{\shortauthors}{Trovato et al.}

\begin{abstract}
  Fuzz testing is one of the most effective techniques for detecting bugs and vulnerabilities in software. However, as the basis of fuzz testing, automated heuristics often fail to uncover deep or complex vulnerabilities. As a result, the performance of fuzz testing remains limited. One promising way to address this limitation is to integrate human expert guidance into the paradigm of fuzz testing. Even though some works have been proposed in this direction, there is still a lack of a systematic research roadmap for combining Human-in-the-Loop (HITL) and fuzz testing, hindering the potential for further enhancing fuzzing effectiveness. To bridge this gap, this paper outlines a forward-looking research roadmap for HITL for fuzz testing. Specifically, we highlight the promise of visualization techniques for interpretable fuzzing processes, as well as on-the-fly interventions that enable experts to guide fuzzing toward hard-to-reach program behaviors. Moreover, the rise of Large Language Models (LLMs) introduces new opportunities and challenges, raising questions about how humans can efficiently provide actionable knowledge, how expert meta-knowledge can be leveraged, and what roles humans should play in the intelligent fuzzing loop with LLMs. To address these questions, we survey existing work on HITL fuzz testing and propose a research agenda emphasizing future opportunities in (1) human monitoring, (2) human steering, and (3) human–LLM collaboration. We call for a paradigm shift toward interactive, human-guided fuzzing systems that integrate expert insight with AI-powered automation in the next-generation fuzzing ecosystem.

\end{abstract}



\maketitle

\section{Introduction}

As software systems continue to grow in scale and complexity, ensuring their security and reliability has become a central challenge in software engineering. Among dynamic testing methods, fuzz testing (fuzzing) has emerged as one of the most effective techniques. By automatically generating and executing large volumes of randomized or mutated inputs, fuzzers are capable of uncovering critical defects, including memory safety violations, boundary errors, and protocol inconsistencies. Since its introduction in the early 1990s~\cite{miller1990empirical}, fuzzing has evolved from simple random testing into a rich ecosystem of tools and techniques, widely deployed in both industry~\cite{afl,bamohabbat2023honggfuzz+} and academia~\cite{bohme2016coverage,chen2018angora}. Fuzzing's ability to explore deep code paths and produce diverse test cases has made it indispensable in software security.

Despite these advances, fuzzing still faces non-negligible limitations. Automated fuzzers often struggle to overcome semantic roadblocks, construct robust drivers, or interpret complex input structures. The large volumes of runtime data they generate are difficult to interpret without domain expertise, and automated heuristics can easily plateau in coverage. These challenges have motivated the integration of human expertise into the fuzzing loop, where analysts can inject constraints, adjust instrumentation, curate high-value seeds, or interpret coverage plateaus. Existing surveys have begun to discuss these issues: Kadron \etal~\cite{kadron2023fuzzing} discuss how expert guidance can augment fuzzing and symbolic execution; Yan \etal~\cite{yan2022survey} categorizes human contributions such as knowledge injection, seed management, and visualization; B{\"o}hme~\cite{bohme2020fuzzing} raises fundamental questions about usability and human in the loop (HITL) communication; and Huang \etal~\cite{huang2025challenges} reviews emerging LLM-for-fuzzing techniques while calling for HITL perspectives. However, these surveys remain fragmented and do not yet provide a systematic phase-oriented taxonomy or consider the evolving interplay between humans and new techniques.


Recent progress in large language models (LLMs) has opened new possibilities for fuzzing. LLMs are capable of synthesizing structured inputs, inferring protocols from documentation, generating fuzz drivers, or even acting as lightweight solvers for path conditions. Early works demonstrate that LLMs can, in some scenarios, partially substitute human roles in fuzzing by providing knowledge, adaptivity, and generative power at scale~\cite{xia2024fuzz4all,deng2023large,meng2024large,shi2025cgifuzz}. Nonetheless, LLMs remain hampered by serious limitations~\cite{huang2025challenges,yang2025hybrid}: they may produce brittle, low-quality, or hallucinated artifacts, struggle with complex reasoning transparency, and suffer from context length limits, inconsistency, or insufficient domain grounding. These defects can lead to invalid test inputs, false positives, or coverage gaps in fuzzing contexts. For instance, recent surveys~\cite{jiang2024fuzzing,huang2024large} identify that existing LLM-based fuzzing works encounter pitfalls such as invalid driver generation, limited input diversity, or model hallucination in reasoning. The emerging paradigm is therefore not human versus LLM, but human–LLM collaboration: future fuzzing workflows will likely blend human oversight, domain judgment, and validation with the generative and adaptive capabilities of LLMs, aiming to balance efficiency and reliability.

Building on insights from the human–computer interaction (HCI) community and motivated by these developments, this paper conducts the first systematic survey that bridges fuzzing, HCI, and LLM-assisted fuzzing. From an initial corpus of 115 papers, we carefully analyzed and identified 44 works that explicitly integrate human involvement into fuzzing. Our analysis reveals that these approaches can be broadly categorized into human monitoring, where humans passively observe and interpret the fuzzing state, and human steering, where humans actively intervene in the fuzzing process to influence exploration.

Our survey reveals that while HITL fuzzing approaches have demonstrated clear benefits, they also require substantial improvement. \fish{detail current limitation, and } Moreover, existing HITL for software testing paradigms still face fundamental limitations, including the high cost of human labor, the limited scalability to large-scale applications, and the lack of generalization across different testers and domains. These constraints highlight the need for a new paradigm that more effectively leverages human expertise while reducing dependence on costly manual intervention. In this context, the rapid rise of LLMs offers a compelling opportunity, which lets LLMs act as collaborators that complement human intuition with scalable automation, opening the path toward human–LLM collaborative fuzzing. Building on this review, we articulate the opportunities and risks of such a paradigm and propose a roadmap toward a sustainable, scalable, and effective fuzzing ecosystem. Our contributions are as follows:
\begin{itemize}
    \item \textbf{Systematic literature review.} We collect and analyze 44 research papers that explore HITL in fuzzing, providing a structured taxonomy of approaches, their strengths, limitations, and application scenarios.
    \item \textbf{Interaction design space analysis.} Drawing on fuzzing's five canonical phases, HCI principles, and models of LLM-assisted fuzzing, we map the design space of human involvement, discussing in the literature review on trade-offs they entail, and what challenges remain unresolved.
    \item \textbf{Visionary HITL fuzzing roadmap.} We extend the discussion to the role of LLMs in fuzzing, highlighting how LLMs can augment or replace human tasks, and we outline research directions about how humans can be involved in future fuzz testing and LLM-assisted fuzzing.
\end{itemize}

The remainder of this paper is organized as follows. Section 2 presents the background of fuzz testing, fuzzing phases, existing HITL paradigms, and LLM-Assisted fuzzing. Section 3 explains the methodology of our literature review. Section 4 reports our detailed literature review on HITL fuzzing across the five phases of fuzzing. Section 5 discusses future research opportunities in HITL and human–LLM collaborative fuzzing, and Section 7 concludes the paper.
\section{Background}

\subsection{Fuzz Testing}
\label{sec:fuzzing}

\begin{figure*}[t]
    \centering
    \includegraphics[width=1\linewidth]{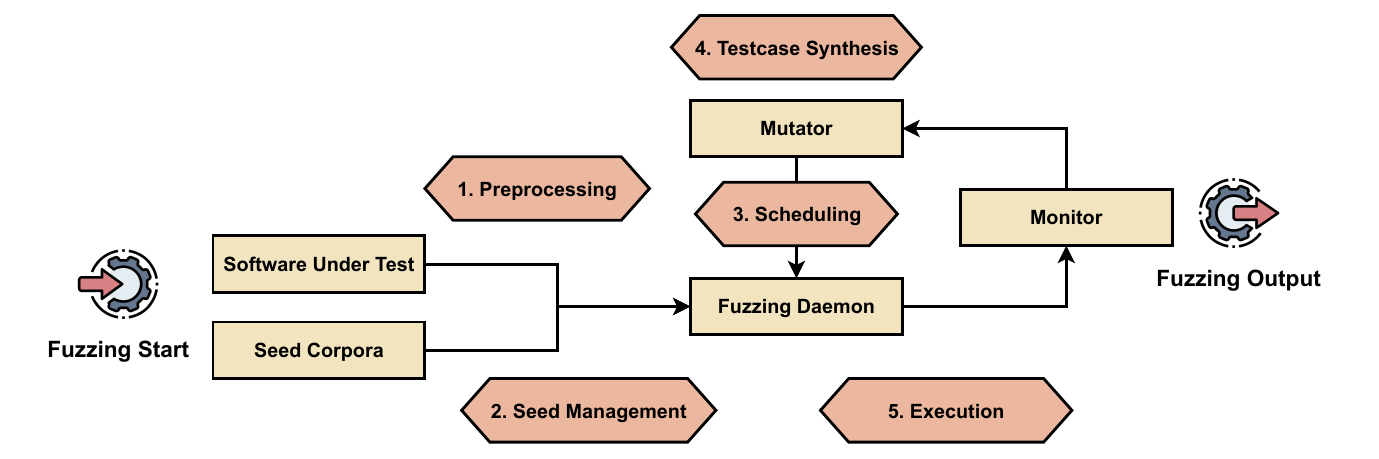}
    \Description{A typical workflow with five key phases for fuzz testing.}
    \caption{A typical workflow of five key phases for fuzz testing.}
    \label{fig:fuzzing}
\end{figure*}

Contemporary fuzzing workflows can be conceptualized as an iterative loop of five interdependent phases~\cite{yan2022survey}, as illustrated in~\autoref{fig:fuzzing}. Specifically:

\begin{itemize}
\item \textbf{Preprocessing}: Preparing the target program for fuzzing by embedding feedback mechanisms (e.g., coverage tracking) through instrumentation or static analysis. This step enables feedback-guided exploration by providing the fuzzer with quantitative signals of execution progress, thereby facilitating more directed input generation.

\item \textbf{Seed Management}: Curating and maintaining the initial pool of test inputs that bootstrap meaningful exploration of the input space. This phase often includes seed minimization and deduplication to ensure that the corpus remains compact yet behaviorally diverse, reducing redundant exploration while preserving coverage potential. Moreover, a high-quality fuzzing corpus can play a decisive role in overcoming otherwise intractable conditional branches and exploration plateaus, thereby enhancing the effectiveness of the entire fuzzing campaign.

\item \textbf{Scheduling}: Prioritizing seeds with the greatest potential to expose novel program behaviors. This is typically achieved through search heuristics and power scheduling strategies that balance efficiency with exploration, ensuring that limited computational resources are allocated to the most promising inputs. Effective scheduling is particularly critical given the vast input space and the inherently constrained testing budget in real-world fuzzing scenarios.

\item \textbf{Testcase Synthesis}: Expanding the input corpus by generating new test cases, either through mutation-based techniques that alter existing seeds or through generation-based approaches that synthesize entirely new inputs. This step is arguably the most crucial in determining the upper bound of a fuzzer’s capability, as the diversity and quality of synthesized inputs directly influence the extent to which unexplored execution paths can be traversed and deeper vulnerabilities uncovered.

\item \textbf{Execution}: Running synthesized test cases against the target program while continuously monitoring runtime behavior. Any anomalous outcomes, such as crashes, hangs, assertion failures, or other oracle-defined misbehavior, are detected and systematically logged for subsequent triage, debugging, and vulnerability analysis.
\end{itemize}

These components form a tightly coupled feedback loop that enables fuzzers to iteratively refine their input corpus, adapt exploration strategies, and progressively drive execution into deeper and less-tested regions of the program. This automated pipeline has made fuzzing one of the most effective techniques for vulnerability discovery. 

\subsection{HITL Fuzzing}

While automation has driven major advances in the efficiency and scalability of fuzzing, existing fuzzers still fall short in scenarios that require semantic understanding or domain-specific reasoning~\cite{wu2024logos,carvalho2024specbcfuzz}. In practice, automated systems exhibit several inherent limitations:

\begin{itemize}
    \item \textbf{Difficulty in handling semantic constraints.} When inputs must satisfy complex structures such as protocols, trees, state machines, or cross-message consistency rules, coverage-guided or mutation-based heuristics inevitably produce a large number of invalid inputs or fail to progress beyond superficial exploration.
    \item \textbf{Challenges in traversing long or sparse execution paths.} Some vulnerabilities lie behind multiple intermediate states or require specific combinations of conditions. Simple mutation often cannot bypass such bottleneck checks in a single step, leaving deep paths unexplored.
    \item \textbf{Lack of explainability and strategy transparency.} Automated fuzzers are rarely able to reveal why exploration stalls on a particular path or where the search is stuck, making it difficult for users to improve the process.
    \item \textbf{Blind spots in resource allocation.} Given finite computational budgets, fuzzers must trade off between broad exploration and deep investigation. Automated heuristics, however, often make suboptimal allocation decisions, wasting resources or missing critical opportunities.
    \item \textbf{Limited support for human feedback.} Current fuzzing workflows provide little opportunity for testers to intervene the strategy while fuzzing is in progress, constraining the potential benefits of expert guidance.
\end{itemize}

Beyond these limitations, existing research~\cite{bohme2020fuzzing} has emphasized in their methodological reflection that fuzzing research should move toward incorporating HITL components. Whereas most contemporary frameworks treat human experts merely as initiators (e.g., setting up the environment or triaging potential bugs), future directions should enable human auditors to engage during the fuzzing process itself—for instance, identifying roadblocks, adjusting exploration paths, or reconfiguring mutation strategies on the fly.

Reflection of fuzzing workflows as illustrated in~\autoref{sec:fuzzing} and existing limitations, inserting well-designed HCI touchpoints into different phases of the fuzzing loop represents a key opportunity to unlock greater potential. By introducing a HITL perspective, fuzzing can be enhanced in several important ways:

\begin{itemize}
    \item \textbf{Bridging semantic gaps.} Human experts can leverage their understanding of protocol semantics, business logic, or vulnerability patterns to guide mutation toward more meaningful directions, thereby reducing the inefficiency of blind or invalid input generation.

    \item \textbf{Breaking exploration bottlenecks.} When automated exploration stalls at specific branches, checkpoints, or path boundaries, experts can intervene by adjusting strategies, introducing intermediate constraints, or manually crafting transition inputs that help the fuzzer progress further.

    \item \textbf{Enhancing explainability and controllability.} Through interactive visualization and control interfaces, analysts can gain insight into the internal state of fuzzing, enabling them to determine when and how to intervene, redirect, or fine-tune the process.

    \item \textbf{Co-governing resource allocation and strategy.} Under resource constraints, human guidance can complement automated heuristics in making high-level scheduling decisions, such as prioritizing certain sub-paths for deeper exploration or deprecating unpromising seeds.
\end{itemize}

Drawing on insights from the broader HCI research community, we summarize the ways in which humans can participate in fuzzing and categorize into two main forms of involvement, as further elaborated in~\autoref{sec:literature-review}. Specifically:

\begin{itemize}
    \item \textbf{Human monitoring (Passive Monitoring).} Testers act primarily as observers, gaining visibility into the fuzzing process via dashboards, coverage heatmaps, path visualizations, and similar artifacts. In this way, humans do not actively change fuzzing policies during execution. They analyze runtime state and results to identify anomalies, diagnose bottlenecks, and discover blind spots, which is useful for situational awareness, post-hoc analysis, and surfacing cues that indicate where intervention might be beneficial.
    \item \textbf{Human steering (Proactive Steering).} Testers take an active role in directing the fuzzer by adjusting seed priorities, tuning mutation weights, injecting constraints, suppressing unproductive paths, or specifying intermediate mutation strategies. Interventions can range from coarse high-level guidance to fine-grained, runtime decisions at particular control points. This semi-automated, hybrid approach allows human expertise to influence exploration policies and search trajectories directly.
\end{itemize}

Compared to the traditional view that treats humans merely as initial configuration providers and final report reviewers, the HITL paradigm advocates for exposing interaction points within the mid-loop stages of fuzzing. This shift enables expert knowledge to be injected into exploration decisions in real time and improves explainability, responsiveness, and adaptability. 

\subsection{LLM in the Loop Fuzz Testing}

The rapid rise of LLMs has opened a new frontier for fuzzing research. Unlike traditional approaches that rely on fixed heuristics or pre-defined grammars, LLMs can synthesize structured inputs, infer constraints from specifications, and even reason about program behavior at scale. This makes them natural candidates to complement or even substitute for human roles in fuzzing, such as seed generation, driver construction, or guided test synthesis. While their reasoning ability remains limited compared to humans, recent work explores LLMs as "intelligent agents" that provide flexible input generation, adaptive exploration, and domain knowledge injection, akin to HITL designs but automated.

A growing body of studies illustrates diverse ways to integrate LLMs into fuzzing workflows. Some focus on seed and input generation. For example, WhiteFox~\cite{yang2024whitefox} applies a multi-agent LLM framework to compiler fuzzing, summarizing optimization triggers and synthesizing triggering test programs. Fuzz4All~\cite{xia2024fuzz4all} adopts a dual-LLM design, which lets one model distill effective prompts, while another generates or mutates inputs iteratively, demonstrating generalization across multiple programming languages. Benchmarking efforts such as Evaluating LLMs for Enhanced Fuzzing provide systematic comparisons, showing early-coverage gains but also exposing limitations in reliability and adaptivity. TitanFuzz~\cite{deng2023large} takes a more direct approach, treating LLMs themselves as fuzzers that synthesize and mutate test programs for deep-learning libraries, coupled with differential testing oracles.

Other works embed LLMs as adaptive solvers or hybrid components. HyLLfuzz~\cite{meng2024large} leverages LLMs as lightweight concolic engines, translating path conditions into natural language and generating satisfying inputs. G2Fuzz~\cite{zhang2025low} synthesizes input generators for complex formats (images, audio, PDFs) when AFL++ stagnates, combining global exploration by LLMs with local mutation fuzzing. Similarly, ChatAFL~\cite{meng2024large} extracts grammar rules from RFCs, augments seed sequences, and generates new messages at plateaus, effectively acting as an automated protocol expert. More advanced control is achieved in CovRL-Fuzz~\cite{eom2024fuzzing}, which trains LLMs with PPO-based reinforcement signals derived from coverage feedback, allowing mutation strategies to evolve dynamically during execution. A distinct research line explores LLM-based driver generation and repair. Zhang \etal~\cite{zhang2024effective} evaluate LLMs for fuzz driver generation, achieving promising success rates with iterative repair. CKGFuzzer~\cite{xu2025ckgfuzzer} extends this direction by combining LLM agents with code knowledge graphs to plan API combinations, generate and repair drivers, synthesize seeds, and even analyze crashes with CWE-aware reasoning. 

These research works have shown that LLMs can automate tasks that were traditionally expert-intensive at the interface between preprocessing and test execution. Yet, unlike humans, LLMs lack accountability and may generate brittle or hallucinated artifacts, raising new challenges for verification and orchestration. This emerging line of research highlights a key direction: designing fuzzing workflows where humans and LLMs complement each other, leveraging the adaptability of LLMs while preserving the oversight and contextual judgment of human experts.

\section{Review Methodology}

The scope of this survey focuses on research that explicitly integrates human involvement into fuzzing workflows. To be within scope, a paper must (1) directly relate to fuzzing and (2) demonstrate concrete instances of HITL contributions, such as visualization-assisted monitoring, seed management, or domain knowledge injection. Papers that merely mention fuzzing without human interaction, or where human input is limited to final result inspection, are excluded. To capture the recent evolution of automation in testing, we also extend our scope to include works where LLMs play analogous roles to human experts by providing seed generation, driver construction, or guided mutation. In total, our survey covers research papers published between 2007 and 2025, spanning software engineering, security, and human–computer interaction venues.

\begin{figure*}[ht]
    \centering
    \includegraphics[width=1\linewidth]{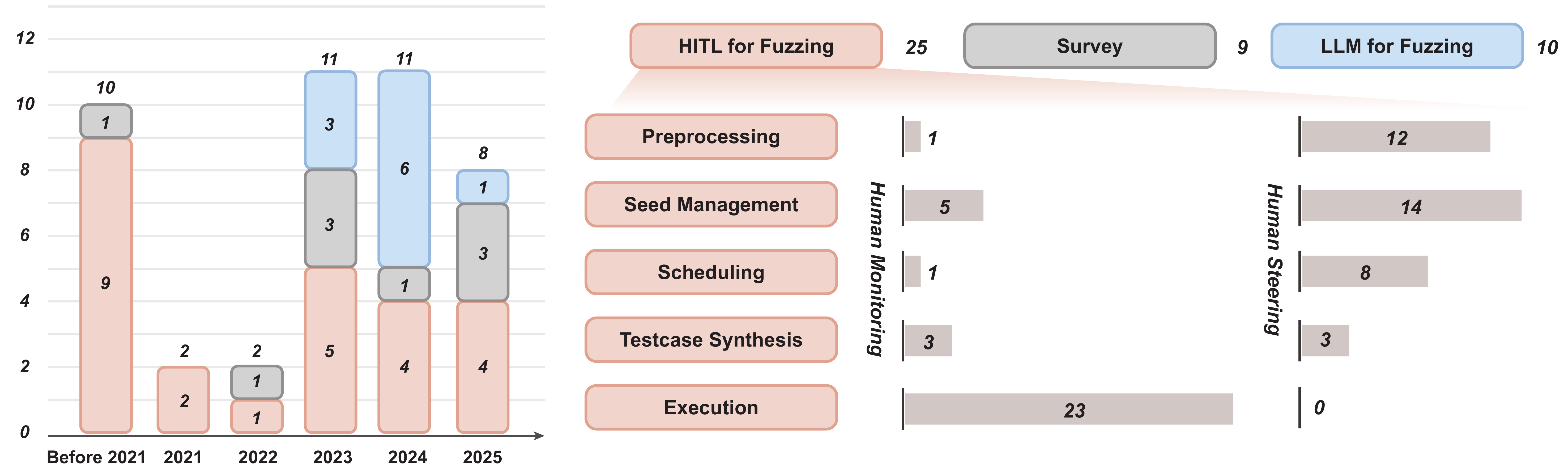}
    \caption{Distribution of collected articles. The left panel shows the temporal trend of published works, grouped by year. The right panel details how articles on HITL for fuzzing are distributed across different fuzzing stages, further distinguishing between human monitoring and human steering roles. Note that the totals may exceed the overall number of reviewed papers, since some works contribute to multiple phases of human monitoring or human steering.}
    \label{fig:statistics}
    \Description{Literature review statistics.}
\end{figure*}


\subsection{Paper Collection}

To provide a comprehensive literature review of existing approaches in conjunction with HITL methods for fuzz testing, we follow established review practices~\cite{zhou2024large,shi2024efficient,zhao2024llm} and design a systematic paper searching to collect relevant references as completely as possible. Our review categorizes prior research according to the five key phases of modern fuzzing frameworks~\cite{kummita2024visualization}, as described in Section~\ref{sec:fuzzing}, and further analyzes them from the two perspectives of human monitoring and human steering in HITL fuzzing. 

First, to ensure broad coverage of the academic landscape, we perform a keyword-based search through Google Scholar as our primary search engine, which provides broad indexing and timely access to published papers and preprints, complemented by targeted searches in IEEE Xplore, ACM Digital Library, and DBLP. This multi-database approach helps ensure a thorough and inclusive collection of relevant papers.



All authors jointly craft three sets of search keywords for paper searching: 

\begin{enumerate}
    \item \textbf{Fuzzing Terms:} This set of keywords establishes the foundational domain of our search. We included terms \textit{fuzzing} and \textit{fuzz test}.
    
    \item \textbf{Human Monitoring Terms:} To capture the mechanisms of human interaction, this set included keywords related to visual interfaces and data representation. The terms used were \textit{visualization}, \textit{visual}, \textit{dashboard}, and \textit{interface}.
    
    \item \textbf{Human Steering Terms:} This set targets the proactive HCI aspect, which is central to our survey. Keywords included were \textit{human}, \textit{interactive}, \textit{guided}, \textit{assisted}, and \textit{interaction}. 
\end{enumerate}

We set the paper date from conferences and journals published over the past twenty years to keep in the recent research for this area, and we engage in discussions to determine their relevance to the HITL fuzzing paradigm. Our criteria for inclusion focused on instances where human intervention directly enhances the fuzzing process at any of these stages, such as modifying mutated seeds or analyzing visualized data to identify bottlenecks. The final queries were constructed by combining fuzzing terms with one or more human visualization terms. Using this method, we retrieved 21 candidate papers among the top 115 results.


\subsection{Snowballing Search}
To further enhance the comprehensiveness of our paper collection, we employed a snowballing search strategy\cite{jalali2012systematic}, a method advocated for its ability to uncover relevant literature that traditional database searches may miss. This process involved two distinct approaches: backward and forward snowballing.
\begin{enumerate}
    \item \textbf{Backward Snowballing:} In this step, we meticulously reviewed the reference lists of the initially identified papers. The aim was to identify foundational or earlier works that are relevant to our survey but were not captured by our keyword-based search. 
    \item \textbf{Forward Snowballing:} Conversely, forward snowballing involves identifying papers that have cited our initial set. This approach allowed us to discover more recent research that builds upon the foundational papers we had already collected. 
\end{enumerate}

Using this snowballing method, we finally found the remaining 23 highly relevant articles.

\subsection{Paper Selection}
\textbf{Inclusion and Exclusion Criteria}. After the papers are collected, we include the following rules based on the discussion in the criteria for whether the papers are selected:

\ding{52} The paper must be in English.

\ding{52} The selected papers must be directly related to fuzzing and explicitly include aspects of human involvement in the fuzzing process.

\ding{55} The paper has fewer than 4 pages.

\ding{55} Books, keynote records, panel summaries, technical reports, tool demos papers, editorials, or venues not subject to a full peer-review process.

\ding{55} Just mentioning fuzzing without human involvement.

\ding{55} Only at the final stage of fuzzing do people examine the results.

We conduct a manual annotation and filtering process to identify papers addressing human involvement in fuzzing. In the first stage, all authors independently read and categorize the candidate papers. We then collectively discuss the categorizations and resolve any discrepancies. In the second stage, we revisit the 12 papers that had been categorized differently by authors to ensure consistency. After careful discussion, we refine our selection to retain only those papers that explicitly involve human participation, while excluding studies focused solely on fully automated fuzzing tools. This secondary screening ensures that the final corpus reflects genuine human-in-the-loop fuzzing research, rather than papers where humans only reviewed final results. Ultimately, we obtain a set of 44 papers, consisting of 35 technical research articles and 9 survey papers, as shown in~\autoref{fig:statistics}.


\section{Current Development}
\label{sec:literature-review}

\begin{table}[htbp]
    \centering
    \small
    \caption{Summary of HITL for fuzzing studies and the phases of the fuzzing workflow where human involvement is introduced. The five phases considered are preprocessing, seed management, scheduling, testcase synthesis, and execution. Human participation is categorized into two modes: \textit{Human Monitoring (HM)} and \textit{Human Steering (HS)}.}
    \label{tab:fuzzing_classification}
    \begin{tabularx}{\textwidth}{X c c c c c c}
        \toprule
        \textbf{} & \textbf{Year} & \textbf{(1) Preproc.} & \textbf{(2) Seed Mng.} & \textbf{(3) Scheduling} & \textbf{(4) Testcase Synth.} & \textbf{(5) Execution} \\
        \midrule
        Drewry et al.~\cite{drewry2007flayer} & 2007 & \HS{HS} & & & & \HM{HM} \\ 
        \midrule
        Machiry et al.~\cite{machiry2013dynodroid} & 2013 &  & \HS{HS} & \HS{HS}& & \HM{HM} \\ 
        \midrule
        Vainio et al.~\cite{vainio2014use} & 2014 & & & & & \HM{HM} \\
        \midrule
        Opmanis et al.~\cite{opmanis2016visualization} & 2016 &  & \HS{HS} & & & \HM{HM} \\    
        \midrule
        Shoshitaishvili et al.~\cite{shoshitaishvili2017rise} & 2017 &  & \HS{HS} & & & \HM{HM} \\
        \midrule
        Cottam et al.~\cite{cottam2017crossing} & 2017 &  & \HS{HS} & & & \HM{HM} \\
        \midrule
        Zhou et al.~\cite{zhou2019visfuzz} & 2019 &  & \HS{HS} & & & \HM{HM} \\
        \midrule
        Aschermann et al.~\cite{aschermann2019nautilus}& 2019 & \HS{HS}  & \HS{HS} & & & \HM{HM} \\
        \midrule
        Aschermann et al.~\cite{aschermann2020ijon} & 2020 & \HS{HS} &  & \HS{HS} & & \HM{HM} \\
        \midrule
        Hussain et al.~\cite{hussain2021fmviz} & 2021 &  & & & \HM{HM} &  \\        
        \midrule
        Fioraldi et al.~\cite{fioraldi2021fuzzsplore} & 2021 &  & & \HS{HS} & \HM{HM} & \HM{HM} \\
        \midrule
        Grishin et al.~\cite{grishin2022human} & 2022 &  & & \HS{HS} & & \HM{HM} \\
        \midrule
        Lu et al.~\cite{lu2023fuzz} & 2023 & & \HM{HM} \HS{HS} & & \HS{HS} & \HM{HM} \\
        \midrule
        Yan et al.~\cite{yan2023infuzz} & 2023 & \HS{HS} & \HM{HM} \HS{HS} & & & \HM{HM} \\
        \midrule
        Bundt et al.~\cite{bundt2023homo} & 2023 & \HS{HS} & \HS{HS} & & & \HM{HM} \\
        \midrule
        Coppa et al.~\cite{coppa2023FuzzPlanner} & 2023 & \HM{HM} \HS{HS} & \HM{HM} \HS{HS} & & &  \HM{HM} \\
        \midrule
        Koffi et al.~\cite{koffi2023efficient} & 2023 & \HS{HS} & \HS{HS} & & &  \HM{HM} \\
        \midrule
        Fang et al.~\cite{fang2024ddgf} & 2024 &  &  \HS{HS} & & &  \HM{HM} \\
        \midrule
        Xu et al.~\cite{xu2024vischeduler} & 2024 &  &   \HM{HM}  &  \HM{HM} \HS{HS} & &  \\
        \midrule
        Gridin et al.~\cite{gridin2024point} & 2024 & \HS{HS}  &   & & \HS{HS}  &   \HM{HM}  \\
        \midrule
        Chambers et al.~\cite{chambers2024hifuzz} & 2024 & \HS{HS}  &  \HS{HS}  &  \HS{HS}  &  &   \HM{HM}  \\
        \midrule
        Li et al.~\cite{li2025vdfuzz} & 2025 &    \HS{HS} &  &   \HS{HS} & & \HM{HM} \\
        \midrule
        Li et al.~\cite{li2025gdfuzz} & 2025 &     &  &   \HS{HS} & \HM{HM} \HS{HS} & \HM{HM} \\
        \midrule        
        Wen et al.~\cite{wen2025prettismart} & 2025 &  \HS{HS} & \HM{HM}   & &  &   \HM{HM}  \\
        \midrule
        Koffi et al.~\cite{koffi2025speeding} & 2025 &  \HS{HS} & \HS{HS} &    & & \HM{HM} \\
        \bottomrule
    \end{tabularx}
\end{table}

To organize our discussion, we structure the surveyed works according to the five canonical phases of fuzzing as outlined in~\autoref{sec:fuzzing}. For each phase, we examine how HITL approaches have been applied, distinguishing between human monitoring, where users passively observe and interpret fuzzing states, and human interactive steering, where users actively intervene to influence exploration. Specifically, from~\autoref{fig:statistics}, we can find that most existing monitoring efforts are concentrated in the execution stage, where visualizations and dashboards are used to observe runtime states under different configurations and to provide testers with actionable feedback about fuzzing progress. By contrast, interactive steering is largely concentrated on seed preparation, such as curating, annotating, or injecting specialized seeds before fuzzing begins, with virtually no work offering real-time HCI interfaces for steering during fuzzing execution. We find that the overall number of HITL for fuzzing studies remains relatively limited, and that this line of work has declined further with the recent rise of LLM-assisted fuzzing, where LLMs increasingly substitute for roles previously envisioned for human experts.

In particular, we conduct a detailed examination of existing HITL papers across the five different phases of the fuzzing workflow as shown in Table~\ref{tab:fuzzing_classification}.

\subsection{Phase \RomanNum{1}: Preprocessing}

Preprocessing lays the groundwork for effective fuzzing by preparing the program under test (SUT) and its environment before input generation and execution commence. It involves instrumentation for coverage-guided fuzzing or taint feedback analysis, the creation of harnesses or drivers, the specification of configuration parameters, and the incorporation of domain knowledge into the fuzzing pipeline. Well-designed preprocessing is essential for reducing invalid inputs, surfacing informative signals, and ensuring that fuzzing explores meaningful program behaviors. However, this phase is notoriously required for expert knowledge as writing robust drivers, building input models, and instrumentation often requires deep domain expertise. From a HITL perspective, preprocessing is a high-leverage phase where even small expert interventions, such as annotations, constraints, or target selections, can reshape the fuzzing landscape. Thus, it provides both significant challenges and promising opportunities for integrating human monitoring and steering.

Existing works have been proposed to assist with the monitoring side. FuzzPlanner~\cite{coppa2023FuzzPlanner} visualizes static and dynamic analysis outputs—including extracted binaries, CVE mappings, and process interactions to enable analysts to better observe the preprocessing pipeline before launching firmware fuzzing campaigns. Similarly, the survey of Qiao \etal~\cite{qiao2025human} highlights that preprocessing in this domain often entails heavy manual setup, such as crafting invariants, initializing blockchain states, or writing harnesses, underscoring the substantial human burden involved in preparing effective fuzzing environments.

More existing research works have been proposed on how humans actively steer preprocessing to overcome roadblocks. InFuzz~\cite{yan2023infuzz} allows testers to annotate code with modified drivers, thereby exposing hidden paths and influencing downstream fuzzing. FuzzPlanner~\cite{coppa2023FuzzPlanner} itself also supports interactive steering, enabling users to select binaries, filter by vulnerability score, and configure experiment parameters to prioritize high-risk components. VDFuzz~\cite{li2025vdfuzz} provides visual interfaces for CFG inspection, allowing analysts to annotate unexplored regions as new targets before recompilation. Other directed fuzzing approaches rely on human-provided guidance. Specifically, analysts explicitly specify target functions or code regions to accelerate symbolic execution toward security-critical paths~\cite{koffi2025speeding}. In cryptographic testing, existing research~\cite{gridin2024point} lets experts inject domain-specific restrictions into test generators, reshaping the input model to bypass cryptographic barriers.

Further, several tools show how minimal preprocessing edits can unlock coverage plateaus. Bundt \etal~\cite{bundt2023homo} suggest where humans should modify harnesses or environment options to expose otherwise unreachable paths. IJON~\cite{aschermann2020ijon} formalizes lightweight annotation primitives that testers insert into source code to expose hidden state or optimization objectives, effectively redefining the fuzzer's feedback function. Flayer~\cite{drewry2007flayer} allows testers to force branch outcomes or skip function calls, reconfiguring program behavior at preprocessing time to bypass guard conditions. 

Domain-specific studies illustrate similar steering roles. In HIFuzz~\cite{chambers2024hifuzz}, testers define the fuzzable task space, including roles, devices, mission sequences, and environmental parameters, at the outset of UAV testing, constraining the search space for meaningful exploration. NAUTILUS~\cite{aschermann2019nautilus} shows that enriching grammars with identifiers or symbol lists, often manually extracted from documentation, significantly improves input generation quality in grammar-based fuzzing. PrettiSmart~\cite{wen2025prettismart} allows users to configure fuzzing parameters such as the number of simulated users, owner addresses, and balance constraints, thereby setting the simulation boundaries before fuzzing begins.

\subsection{Phase \RomanNum{2}: Seed Management}

Seed Management is concerned with curating, prioritizing, and evolving the pool of test inputs that drive a fuzzer's exploration. In this phase, tasks include selecting initial seeds, deduplicating, clustering or classifying them, and injecting new seeds to guide coverage growth. Good seed management helps avoid wasted cycles on redundant or low-value inputs while promoting diversity and reaching underexplored program regions. For HITL designs, Seed Management is especially attractive as human experts can help monitor the seed-to-path relationships, spot coverage gaps, and inject or filter seeds based on domain insight.

Existing research on the monitoring side, FuzzInspector~\cite{lu2023fuzz} highlights seed–path differences and low-coverage functions, aiding users in observing the evolution of the seed corpus and identifying under-tested inputs. InFuzz~\cite{yan2023infuzz} displays which seeds reach near bottleneck locations, enabling testers to inspect input–path relationships and detect gaps in exploration. FuzzPlanner~\cite{coppa2023FuzzPlanner} allows users to visualize which communication channels or seed sources are yielding valuable inputs, helping select promising channels during seed collection. ViScheduler~\cite{lu2023fuzz} uses embedding-based visualizations of seed distributions, allowing experts to monitor clustering, anomalies, and coverage diversity in the seed pool. Moreover, Qiao \etal~\cite{qiao2025human} emphasizes that the way practitioners curate seeds or define target sequences at the outset significantly affects exploration results, and PrettiSmart~\cite{wen2025prettismart} visualizes fuzz-generated input sequences, helping human users observe bias, diversity, and coverage dynamics in the seed space.

On the steering side, FuzzInspector~\cite{lu2023fuzz} allows users to define seed constraints, effectively steering the fuzzer toward particular branches via seed filtering or restriction. Further works~\cite{yan2023infuzz,koffi2023efficient} let testers add new seeds to bypass blocking branches found via visualization. Real user input streams can also be captured and reused as seeds~\cite{cottam2017crossing}, preserving domain relevance while enabling mutation-based exploration. FuzzPlanner~\cite{coppa2023FuzzPlanner} supports user confirmation of chosen seed inputs (e.g., from emulation traces or PoCs) into the corpus, letting humans filter and prioritize seeds. VisFuzz~\cite{zhou2019visfuzz} enables testers to manually construct and inject targeted seeds (such as JSON instances) to break through semantic bottlenecks revealed in visualization. Koffi \etal~\cite{koffi2025speeding} argue that experts can validate and refine the initial seed set using visual feedback, discarding unproductive seeds and updating target lists. HaCRS~\cite{shoshitaishvili2017rise} permits human assistants to propose new seeds by exploring program logs and feeding those into the fuzzing corpus. Recent work has also analyzed~\cite{bundt2023homo} targeted seeds based on ranked ``compartments'' to unlock large unexplored regions. HIFuzz~\cite{chambers2024hifuzz} proposes a layered design that, after clustering L1 results, experts can select representative or boundary tests as seeds to push to L2, refining the seed corpus actively. Dynodroid~\cite{machiry2013dynodroid} gives testers the opportunity to interleave manual meaningful inputs (e.g., login credentials, event sequences) to enrich the seed pool. Moreover, in NAUTILUS~\cite{aschermann2019nautilus}, while generation is automated, users can inspect and select minimized "interesting" seeds forming the evolving seed pool for further splicing and mutation.

\subsection{Phase \RomanNum{3}: Scheduling}

The scheduling phase in fuzzing governs which seed is selected next and how much energy (time/computational quanta) to allocate. A sound scheduler balances exploitation (mutating promising seeds) and exploration (trying less-seen seeds) so that the fuzzer does not waste cycles on fruitless inputs or get stuck on local maxima. In large-scale fuzzing campaigns, scheduling decisions become even more critical, as the scheduling space is vast and dynamic, making it difficult to choose from the huge search space. Human intervention in scheduling is relatively rare, as manually ordering or adjusting vast queues is impractical; nonetheless, experts may write rules or dynamically adjust priority weights or queues in response to observed behavior, offering a pragmatic HITL lever.

Prior works have been proposed to monitor the scheduling in fuzzing, such as ViScheduler~\cite{xu2024vischeduler}, which provides real-time visual feedback about seed cluster distributions, anomalies, and their evolution over time, enabling analysts to observe how scheduling decisions correlate with coverage progress. Such visual insight lets experts spot imbalances or stagnation so they might intervene. While on the steering side, previous works that embed human control in scheduling tend to treat the human as a strategic overseer who adjusts priority rules or responds to observed bottlenecks, rather than micromanaging every seed. Specifically, FuzzSplore~\cite{fioraldi2021fuzzsplore} empowers analysts to reallocate CPU time, drop underperforming fuzzers, or adjust parameter weights mid-campaign, thereby shifting the scheduling strategy on the fly in response to empirical patterns seen in coverage and mutation metrics. Building on introspection, DDGF~\cite{fang2024ddgf} gives users the ability to flag interesting paths during runtime, causing the scheduler to bias toward seeds that exercise those paths more heavily. Similarly, ViScheduler~\cite{xu2024vischeduler} lets experts influence seed prioritization by amplifying energy to select clusters or outliers, altering the scheduler’s preference based on visual cues of seed space structure. In directed fuzzing contexts, VDFuzz~\cite{li2025vdfuzz} allows testers to dynamically add unexplored basic blocks or target points via visualization, which the scheduling component then favors in future seed selection. GDFuzz~\cite{li2025gdfuzz} goes further by combining explanations with a dual-queue balancing algorithm, letting users adjust queue behavior or weights according to insight into which input features matter most. The annotation-based IJON~\cite{aschermann2020ijon} supports runtime toggles between coverage and maximization queues, letting users shift scheduling policy during fuzzing via a simple parameter. Meanwhile, HIFuzz~\cite{chambers2024hifuzz} involves human decisions at higher abstraction levels: its layered design lets experts decide which test batches escalate across layers or real-world execution, effectively serving as a scheduler gatekeeper between modes. Grishin \etal~\cite{grishin2022human} propose that analysts can pause fuzzing, reorder the AFL queue, or switch targeted functions at runtime, injecting manual preference into the scheduling order under limited interaction cost. Even in the hybrid mobile realm, Dynodroid~\cite{machiry2013dynodroid} allows users to pause automatic generation, inject custom events, and alter which seeds the scheduler next prioritizes.

\subsection{Phase \RomanNum{4}: Testcase Synthesis}

Test case synthesis is the key phase for fuzzing, where fuzzers generate new inputs, typically through mutation of existing seeds or generation methods. Most synthesis strategies are automated and encoded in advance, e.g., bit/byte flips in AFL~\cite{fioraldi2020afl++}, leaving little room for direct human involvement. However, fuzzing is inherently dynamic: the utility of particular mutation operators may change as execution progresses, and static mutation heuristics may fail to adapt. From an HITL perspective, this phase presents opportunities for experts to monitor mutation patterns in real time and, in limited cases, to steer mutation by injecting domain constraints or selectively amplifying features. 

Existing work on monitoring the testcase synthesis aims to help analysts make sense of how test cases are generated. FMViz~\cite{hussain2021fmviz} visualizes AFL-generated inputs at the byte level, allowing users to see mutation patterns across generated test cases and track the fuzzer’s behavior as inputs evolve. FuzzSplore~\cite{fioraldi2021fuzzsplore} presents test case clusters and mutation trajectories, helping users observe whether the fuzzer is exploring diverse input regions or converging too narrowly. GDFuzz~\cite{li2025gdfuzz} complements this by showing byte-level feature importance maps derived from explainable AI, giving testers visibility into which parts of inputs matter most for reaching target paths. 

On the expert steering side, existing works demonstrate how humans can actively shape test case generation. FuzzInspector~\cite{lu2023fuzz} enables testers to inject constraints derived from seed–CPU state analysis, effectively steering new test case generation toward unexplored states. GDFuzz~\cite{li2025gdfuzz} extends its monitoring interface into steering by letting users refine sample masks, selectively protecting or exposing bytes to guide mutation strategies toward meaningful variations. In a domain-specific setting, Gridin \etal~\cite{gridin2024point} show how human operators can insert cryptographic constraints into a test generator, so the generated inputs can bypass cryptographic roadblocks and exercise deeper logic.

\subsection{Phase \RomanNum{5}: Execution}

The execution phase is the most frequently targeted in prior research, as it produces outputs that are directly visible to developers and testers. Execution not only generates raw bug signals but also provides continuous feedback that guides the rest of the fuzzing loop. Because execution can yield massive volumes of low-level data, HITL designs often emphasize visualization and monitoring at this phase. It is worth noting, however, that oracle design, which decides whether a runtime outcome constitutes a bug, typically occurs after execution and remains a separate challenge, which is beyond the scope of this survey.

A substantial body of work has focused on visualization in the execution phase. Early efforts such as Vainio~\cite{vainio2014use} introduced web-based dashboards that combined runtime fuzzer data with host metrics, showing crashes over time and multi-host progress to help users quickly assess fuzzing health. Opmanis \etal~\cite{opmanis2016visualization} extended this approach to industrial regression settings, offering matrix, timeline, and pie-chart dashboards for monitoring large campaigns, detecting regressions, and tracing root causes. These studies demonstrate how visualization can transform raw logs into actionable insights.

Other systems enhance execution monitoring by combining real-time coverage profiling with interactive graphics. InFuzz~\cite{yan2023infuzz} reports bottleneck branches and contextual data flow to reveal obstacles to coverage expansion, while VisFuzz~\cite{zhou2019visfuzz} and FuzzSplore~\cite{fioraldi2021fuzzsplore} visualize execution dynamics through charts, call graphs, and mutation trajectories, helping users detect convergence or stalls. Directed fuzzers extend these ideas with finer-grained metrics: DDGF~\cite{fang2024ddgf} and VDFuzz~\cite{li2025vdfuzz} display path profiles, heatmaps, and function hit frequencies to expose critical execution paths.

Domain-specific studies further highlight execution monitoring. PrettiSmart~\cite{wen2025prettismart} visualizes smart contracts via function calls, state updates, and fund transfers, making complex runtime behaviors accessible even to non-experts, while HIFuzz~\cite{chambers2024hifuzz} applies monitoring to cyber-physical testing. Early efforts also explored runtime introspection: Flayer~\cite{drewry2007flayer} surfaces taint flow and conditional jumps so analysts can intervene on stalls, and HaCRS~\cite{shoshitaishvili2017rise} integrates execution traces with a Human–Automation Link, enabling humans to propose new inputs from observed behaviors.

However, among the visualization approaches discussed above, we find no existing work on human steering, specifically for the dynamic alteration of execution strategies during fuzzing. This gap reveals a key research opportunity and motivates one of our future directions.

\subsection{Comparison with Existing Surveys}

Existing surveys have explored the intersection of human expertise and fuzzing, each contributing valuable perspectives to the discourse. Specifically, Kadron \etal~\cite{kadron2023fuzzing} synthesizes expert-in-the-loop levers across fuzzing and symbolic execution, arguing that small, well-placed expert inputs can significantly improve path depth and bug-finding efficiency. Zhang \etal~\cite{zhang2024survey} provides a categorical overview of HITL fuzzing, identifying key areas such as knowledge injection, interactive seed management, and visualization-assisted monitoring. Complementing these, empirical studies conducted by Nourry \etal~\cite{nourry2023human} and Qiao \etal~\cite{qiao2025human} uncover the practical usability barriers and manual burdens faced by developers, emphasizing needs in setup, configuration, and interoperability. 

From a visualization standpoint, Kummita \etal~\cite{kummita2024visualization} offer structured task taxonomies and advocate for frameworks that expose internal fuzzing metrics to support expert monitoring. Reflective research conducted by B{\"o}hme \etal~\cite{bohme2020fuzzing} frames broader HITL questions concerning communication, usability, and dynamic human direction.

While these works collectively underscore the importance of human involvement and begin to map the design space, they remain fragmented in their treatment of the non-comprehensive fuzzing lifecycle (e.g., limited to seed management) or narrowed HITL approaches (e.g, only focus on visualization instead of interactive human steering). Furthermore, in the new era of LLM-assisted testing age, they do not consider the AI-collaborated paradigm of HITL for fuzzing.
\section{The Road Ahead}

To advance next-generation HITL for fuzzing, we identify promising directions for future research. These directions can be organized into three stages, each reflecting a distinct role that humans may assume in the fuzzing process:

\begin{itemize}
\item \textbf{Stage \RomanNum{1}: Towards Fine-Grained and Explainable Fuzzing Monitoring.} 
Develop a unified multimodal fuzzing data pool and observability model, supported by a standardized data schema, to integrate signals across preprocessing, seed management, scheduling, synthesis, and execution. Such a framework should enable cross-phase tracing of fuzzing performance, support comparative analysis across tools, and provide explainable diagnostics of inefficiencies. Future extensions may incorporate LLM-driven reasoning to enhance root-cause analysis and interpretation.

\item \textbf{Stage \RomanNum{2}: Achieving Real-time and Strategic Human Steering.} 
Enable timely and efficient human interventions that guide fuzzing strategies during execution without requiring costly reruns or disrupting other workflow components. This involves designing runtime pipelines that expose actionable information for in-situ analysis and accept human guidance dynamically, together with advanced interaction techniques such as parameterized components, interactive dashboards, what-if analyses, and natural language interfaces. These mechanisms would allow human expertise to be encoded as reusable steering policies that flexibly adapt to different fuzzing contexts.

\item \textbf{Stage \RomanNum{3}: Towards Synergistic Human–LLM Collaboration in Fuzzing.} Leverage LLMs as ``expert assistants'' that augment existing processes by partially substituting for human roles in the fuzzing loop. Afterwards, LLMs can act as autonomous agents or multi-agent collaborators to simulate expert decision-making and low-level cognitive tasks. As LLM capabilities advance, humans transition to high-level commanders, injecting domain expertise and requirements into multi-agent LLM systems to achieve efficient division of labor.

\end{itemize}

\begin{figure*}[ht]
    \centering
    \includegraphics[width=\linewidth]{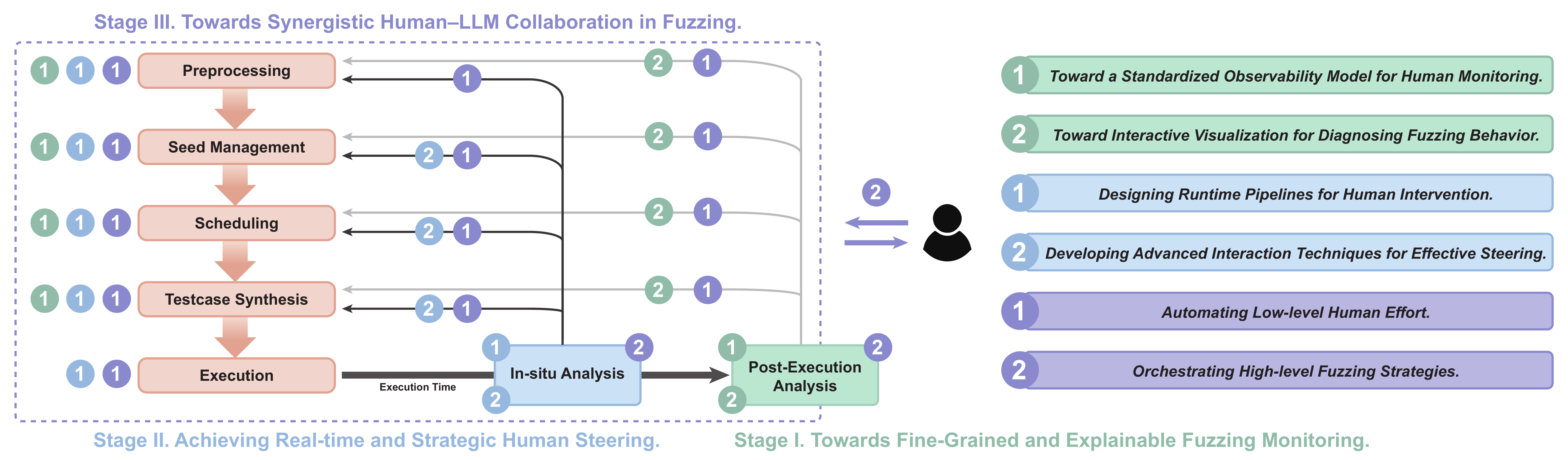}
\caption{The interaction design space for human involvement in fuzzing across three stages. 
Numbered circles mark six future research directions and their corresponding phases in the fuzzing pipeline. 
Gray arrows denote traditional offline feedback after execution, while black arrows denote in-situ runtime intervention. 
Stage~\RomanNum{2} highlights the transition from offline post-execution analysis to real-time supervision and steering.}
    \Description{The interaction design space.}
    \label{fig:future}
\end{figure*}




As illustrated in \autoref{fig:future}, our roadmap conceptualizes human involvement in fuzzing as a progressive evolution across three stages. Stage~\RomanNum{1} emphasizes richer monitoring through standardized observability and interactive visualization. Stage~\RomanNum{2} advances this by establishing runtime pipelines and interaction techniques that move beyond offline analysis toward in-situ steering, serving as a critical bridge between post-execution feedback (gray arrows) and continuous runtime intervention (black arrows). Stage~\RomanNum{3} extends the trajectory further by incorporating LLMs as assistants and orchestrators, ultimately envisioning multi-agent collaboration in which humans transition into supervisory and directive roles. The numbered circle markers indicate six concrete research directions, positioned within the fuzzing pipeline to highlight their potential contributions.

These stages outline a promising research agenda for the evolution of HITL for fuzzing. From our literature review, we observe that progress requires not only greater integration of humans into the loop but also deeper utilization of the fundamental knowledge that human experts possess, as human knowledge cannot be easily replaced or surpassed by current models. Thus, the path forward unfolds in three phases: (1) strengthening HITL designs to enhance traditional fuzzing, (2) employing LLMs to substitute for portions of HITL roles, and (3) enabling synergistic collaboration between humans and advanced LLMs to jointly accomplish fuzzing tasks. The ultimate goal is a human–LLM symbiosis that combines expertise and coordination to achieve scalable, adaptive, and efficient fuzzing workflows.


\subsection{Stage \RomanNum{1}: Towards Fine-Grained and Explainable Fuzzing Monitoring} 

Most current fuzzing workflows still follow an offline, iterative paradigm: testers configure initial parameters such as seeds, scheduling policies, or test-case generators, execute the fuzzer, and then analyze end-of-run results to decide how the configuration should be updated for the next round. Consequently, human insight is largely confined to the post-execution stage. These analyses are typically limited to examining the relationship between seeds, coverage metrics, and bug oracles, while many intermediate processes, such as mutation operators and other runtime artifacts, remain underexplored. As a result, testers gain only partial visibility into how fuzzing strategies evolve during execution and where inefficiencies arise. In addition, current analytical methods remain coarse-grained, relying primarily on control-flow coverage or simple aggregate reporting. Taken together, these limitations constrain the comprehensiveness of human monitoring and highlight the need for richer, multi-faceted approaches that can capture the dynamics of fuzzing beyond coverage and crashes. Future work can proceed along two directions:

\textbf{Direction 1: Towards a Standardized Observability Model for Human Monitoring.}  
To make human monitoring more comprehensive, future research should focus on capturing and analyzing richer signals across all phases of the fuzzing pipeline. From Table~\ref{tab:fuzzing_classification}, it is evident that most instances of Human Monitoring (\HM{HM}) are conducted during fuzzing execution, and that the primary sources of analysis are determined by the instrumentation inserted during the preprocessing phase. However, intermediate aspects such as scheduling and mutation, which significantly influence fuzzing efficiency, are rarely incorporated into the human monitoring process. This narrow focus limits the scope of actionable insights that humans can obtain and constrains the potential benefits of HITL approaches in improving fuzzing strategies.  

Beyond coverage and oracle outputs, intermediate artifacts can provide valuable context for diagnosing inefficiencies and guiding configuration updates. In preprocessing, current workflows often stop at coverage instrumentation, yet static program features such as control-flow complexity, data dependencies, or input parsing routines could be leveraged to anticipate bottlenecks and prioritize instrumentation strategies~\cite{ganesh2009taint,cadar2008klee}. In scheduling, analyses typically rely on aggregate coverage, overlooking finer metrics such as seed selection frequency, execution cost, or incremental coverage contribution, which could reveal biases in power scheduling policies~\cite{bohme2017directed}. Test case synthesis likewise treats mutation operators as black boxes, with little attention to which operators are applied, how often, and under which contexts; recording mutation paths or generation parameters could highlight which strategies consistently yield valuable behaviors and which waste resources~\cite{lemieux2018perffuzz}. Finally, execution and monitoring are usually reduced to crashes or hangs, while richer runtime traces such as memory access patterns, system call sequences, or taint propagation paths are rarely analyzed.  

To support such a comprehensive analysis, we envision building a unified multimodal fuzzing data pool and observability model, supported by a standardized data schema. Such a model would systematically integrate signals from preprocessing, seed management, scheduling, test case synthesis, and execution, enabling comparative analysis across different fuzzing tools and experiments. By correlating heterogeneous artifacts, for instance, linking seed provenance to scheduling priorities, mutation operators to coverage growth, and execution results to bug discovery, this unified framework would not only enhance interpretability but also establish a common foundation for benchmarking and cross-tool comparison.

\textbf{Direction 2: Toward Interactive Visualization for Diagnosing Fuzzing Behavior.}  
A recent study~\cite{kummita2024visualization} developed a task taxonomy for understanding fuzzing internals based on interviews with professional fuzzing testers. Their findings highlight that testers often need to explore complex relationships, make comparisons, and track temporal dependencies across fuzzing artifacts, yet most existing visualization tools provide little support for these demands. Current analyses in fuzzing remain largely restricted to aggregate coverage rates or coarse metrics such as control-flow and function-call coverage~\cite{zhou2019visfuzz, fioraldi2021fuzzsplore}. While these measures provide a basic sense of progress, they fail to support multi-perspective or fine-grained comparisons that are critical for uncovering inefficiencies in fuzzing behavior~\cite{gleicher2017considerations}. Building on the richer data sources discussed in Direction~1, future research should therefore focus on developing interactive visualization systems tailored to fuzzing, with a particular emphasis on comparative and diagnostic analysis. Such systems could help testers not only assess overall effectiveness but also trace the causes of inefficiencies and derive actionable strategies for improvement.  
Yet existing fuzzing visualization approaches rarely support such fine-grained, multi-level exploration, 
instead of remaining limited to aggregate metrics or single-layer comparisons. 
Similar challenges have been addressed in other domains through behavior visualization systems 
that capture hierarchical processes and multi-perspective interactions~\cite{wen2024ponzilens+}, 
suggesting possible design directions for fuzzing.
In scheduling, visual analytics might reveal whether power scheduling policies disproportionately favor low-cost seeds while neglecting rare but high-value inputs. In seed management, dashboards could expose redundancy or diversity in seed pools and illustrate their downstream impact on synthesis efficiency and coverage expansion. By presenting these perspectives in an interpretable and interactive form, visualization systems would enable humans to diagnose hidden inefficiencies, explore ``what-if'' alternatives~\cite{wexler2019if}, and intervene more effectively during fuzzing.  

The key challenge lies in balancing depth and scalability: fuzzing generates large volumes of heterogeneous, rapidly evolving data, and naive visual encodings risk overwhelming users. Future work must therefore investigate visualization designs that scale to large input corpora, support temporal and structural comparisons across fuzzing phases, and provide abstractions that are both expressive and cognitively manageable. Addressing these challenges would position visualization as a central mechanism for empowering human testers to guide and refine fuzzing strategies in practice.

\textbf{Summary.} Stage \RomanNum{1} emphasizes moving beyond coarse, outcome-oriented metrics toward fine-grained and explainable monitoring of fuzzing processes. By developing a standardized observability model that integrates heterogeneous signals across all phases and by designing interactive visualization systems for comparative and diagnostic analysis, humans can obtain multi-level, multi-perspective insights into fuzzing behavior. Such comprehensive monitoring enables more accurate identification of inefficiencies, supports evidence-based configuration updates, and lays the analytical foundation for Stage \RomanNum{2}, where these insights can be leveraged for real-time steering and intervention within the fuzzing loop.

\subsection{Stage \RomanNum{2}: Achieving Real-time and Strategic Human Steering}
From Table~\ref{tab:fuzzing_classification}, it is evident that existing efforts on Human Steering (\HS{HS}) remain highly uneven across the fuzzing pipeline. Most interventions are concentrated in early stages, such as preprocessing and seed management, where humans configure instrumentation or curate input corpora. In contrast, scheduling and test case synthesis are still dominated by automated processes in state-of-the-art fuzzers~\cite{nourry2023human}, with limited transparency or opportunities for human guidance. This imbalance suggests significant opportunities for research on making these intermediate stages more interpretable and steerable. Moreover, most HS instances occur offline rather than as in-situ, runtime guidance, leaving steering decisions decoupled from the actual execution loop.  
As discussed in Stage \uppercase\expandafter{\romannumeral 1}, this offline paradigm is often time-consuming and inefficient, since every adjustment requires rerunning the fuzzer from scratch. A promising direction is therefore to enable real-time analysis and timely intervention, allowing humans to guide exploration toward critical code regions, avoid wasting resources on low-value paths, and improve overall efficiency. In essence, Stage \uppercase\expandafter{\romannumeral 2} extends the advances in post-execution analysis from Stage \uppercase\expandafter{\romannumeral 1} into runtime, so that richer insights can directly inform steering decisions while the fuzzer is still running. Specifically, possible research directions include:

\textbf{Direction 1: Designing Runtime Pipelines for Human Intervention.}  
Most fuzzing workflows are highly automated and rigid, offering little opportunity for humans to intervene once execution begins. A promising research direction is to design runtime pipelines that support in-situ human intervention without requiring repeated restarts, transforming fuzzing from a closed-loop automation into a collaborative process. Such pipelines should exhibit three essential features.
First, they must be capable of collecting and exposing sufficient runtime information to support meaningful analysis by human testers. Building on the richer monitoring signals discussed in Stage~\RomanNum{1}, this could include exposing per-seed contribution to coverage, mutation lineage, or operator effectiveness in real time. For instance, schedulers could display incremental coverage contributions of seeds as they are executed, enabling testers to detect when the fuzzer is over-exploiting shallow seeds or failing to invest in deeper, high-potential ones. Similarly, mutation operators could expose success rates or coverage yields as streaming metrics, making it possible to identify ineffective operators during long fuzzing runs~\cite{zhao2022amsfuzz}.
Second, the pipeline should enable humans to convey their guidance back to the fuzzer in situ, without disrupting ongoing execution. Instead of restarting the fuzzer with a new configuration, testers should be able to adjust steering policies dynamically—for example, raising the weight of promising mutation operators, deprioritizing redundant seeds, or temporarily focusing exploration on security-critical modules. This raises open questions about what forms of guidance are most natural and effective, whether numeric parameter adjustments, high-level constraints, or rule-based directives.
Third, the fuzzer must incorporate human input efficiently, adapting its execution logic without undermining scalability. One possible approach is to design modular fuzzing loops where scheduling, mutation, and triaging are parameterized components that can be updated at runtime. For example, a scheduler could accept human-specified priority hints as an overlay on top of automated heuristics, while mutation engines could dynamically swap operator profiles based on user preferences~\cite{luo2024make}. These designs would preserve throughput while enabling responsiveness to human steering.

Together, these methods allow fuzzing to evolve from a rigid automation into an interactive workflow where humans can continuously diagnose inefficiencies, test alternative strategies toward more valuable execution paths.

\textbf{Direction 2: Developing Advanced Interaction Techniques for Effective Steering.}  
Once pipelines are able to capture runtime information and accept user guidance, the challenge shifts to designing effective interaction techniques that allow humans to express their intent in a usable and efficient manner. Current practice often requires modifying the fuzzer’s source code or internal logic directly, which demands substantial expertise and is both time-consuming and error-prone. A promising direction is to organize the fuzzing workflow as a unified framework composed of parameterized components, where each stage, such as seed management, scheduling, or mutation, can be adjusted through configurable parameters rather than code changes. This idea parallels modular visualization and workflow systems in other domains, such as VisFlow~\cite{mou2017visflow} and Vega-lite parameterization~\cite{satyanarayan2016vega}, which lower the barrier for analysts by exposing pipelines as configurable, declarative components.

In the longer term, more flexible paradigms could be developed, such as interactive visual interfaces that expose the fuzzing pipeline as a composition of modules on a canvas. Here, users could drag and configure components, adjust high-level fuzzing strategies, and immediately observe the implications of their choices, similar to visual programming environments like Orange~\cite{demvsar2013orange} or Envisage~\cite{wen2025envisage}. What-if analyses could further augment this process by simulating the potential outcomes of alternative strategies, drawing inspiration from decision-support tools in visual analytics~\cite{badam2016timefork}. For instance, a fuzzing dashboard could allow testers to explore how different scheduling policies might affect long-running server programs versus lightweight utilities, or how alternative mutation strategies perform on software with complex input formats compared to those with simple command-line arguments.

By enabling such scenario-specific exploration, humans can tailor fuzzing strategies to diverse software contexts, improving efficiency and effectiveness. Beyond visual manipulation, other modalities could also be explored. Natural language interfaces, which have shown promise in data analysis systems~\cite{setlur2016eviza}, could allow testers to issue steering commands such as ``increase focus on parsing routines'' without needing low-level configuration. Together, these approaches suggest that HCI methods already successful in other interactive domains can be adapted to make fuzzing steering more usable, flexible, and effective.

\textbf{Summary.}  
Stage \uppercase\expandafter{\romannumeral 2} highlights the transition from post-execution adjustments to real-time, strategic steering of fuzzing. By establishing runtime pipelines that expose actionable signals and accept timely human input, testers are no longer confined to slow offline iterations but can directly influence ongoing executions. In parallel, advanced interaction techniques such as parameterized workflows, interactive what-if dashboards, and natural language interfaces reduce the barrier for expressing intent and enable humans to guide fuzzing at both fine-grained and high-level stages. Together, these directions position human steering not as a reactive configuration but as a proactive and integrated component of the fuzzing loop, ensuring that computational resources remain aligned with expert knowledge and goals.

\subsection{Stage \RomanNum{3}: Towards Synergistic Human–LLM Collaboration in Fuzzing}
Recent advances in large language models (LLMs) have demonstrated strong capabilities in program comprehension, semantic reasoning, natural language interaction, and adaptive decision-making. These advances create new opportunities to extend the role of humans in fuzzing beyond offline analysis and runtime steering toward synergistic collaboration with intelligent systems. We envision this integration along two directions.  

\textbf{Direction 1: Automating Low-level Human Effort.}  
A significant portion of human involvement in fuzzing today is consumed by repetitive and mechanical tasks, such as deduplicating seeds, tuning scheduling parameters, or filtering false positives in crash reports. Recent LLM-based fuzzing research demonstrates that these activities can increasingly be automated. For instance, LLMs have been shown to generate diverse seeds and mutations through iterative prompting and autoprompting loops~\cite{xia2024fuzz4all,deng2023large}, synthesize generators for complex non-textual formats to replace manual input engineering~\cite{zhang2025low}, and automatically extract or repair protocol grammars for structured fuzzing~\cite{meng2024large}. Similarly, LLMs can aid in crash analysis and driver generation: recent work shows their effectiveness in automatically producing fuzz drivers, clustering crash reports, and refining test harnesses that would otherwise require significant human effort~\cite{zhang2024effective,xu2025ckgfuzzer}.  

Building on these advances, future work could extend the use of LLMs to runtime support, such as reasoning over execution traces to prune redundant seeds, recommending parameter adjustments based on coverage growth, or categorizing crashes into unique vulnerability classes. The main challenge lies in balancing automation with reliability: LLM-generated adjustments must not introduce bias, instability, or missed opportunities, particularly in long-running fuzzing campaigns where subtle errors can accumulate. Another key issue is transparency: LLMs must expose their reasoning process, for example, through interpretable decision logs or explainable ranking outputs, so that testers can validate recommendations before adoption. Successfully addressing these challenges would free humans from mechanical oversight and allow them to focus on higher-level strategic decisions.  

\textbf{Direction 2: Orchestrating High-level Fuzzing Strategies.}  
Beyond automating low-level tasks, LLMs have the potential to assume responsibilities traditionally reserved for experienced testers~\cite{wang2025leveraging}, such as designing fuzzing plans, evaluating ongoing effectiveness, and adapting strategies to different software domains. Recent systems illustrate this possibility: LLMs have been used as lightweight symbolic solvers to augment hybrid fuzzing~\cite{meng2024large2}, to generate high-level fuzzing plans for compiler testing~\cite{yang2024whitefox}, and to orchestrate protocol-specific fuzzing through adaptive plateau-breaking inputs~\cite{meng2024large}. More generally, frameworks such as ~\cite{black2024evaluating} have begun to establish evaluation methodologies for LLM-driven fuzzing strategies, providing a foundation for benchmarking orchestration roles.  

Such orchestration requires LLMs to continuously monitor runtime metrics and human-provided goals, then translate these heterogeneous signals into coordinated fuzzing actions. For example, an LLM could propose a high-level strategy that balances exploration and exploitation, dynamically decide when to switch between grammar-based and mutation-based fuzzing, or assign specialized modules to protocol fuzzing versus command-line utilities. Compared with Stage~\RomanNum{1}, where humans perform post-execution analysis, or Stage~\RomanNum{2}, where humans intervene during runtime, LLM-augmented fuzzing envisions a more autonomous process where humans shift into the role of supervisors and directors.  

A particularly promising line of research is the design of \textit{multi-agent LLM fuzzing systems}, where distinct LLM agents emulate human roles at scale. For example, Explorer Agents could specialize in generating input variations, Analyst Agents could triage crashes by leveraging vulnerability knowledge graphs, and Strategist Agents could optimize scheduling policies using historical campaign data. Natural language interfaces could allow humans to issue high-level directives, which the system would translate into policy updates or reward function adjustments. By distributing responsibilities across specialized LLM agents, such systems would reduce the cognitive load on testers while enabling more adaptive and scalable orchestration.  

Achieving this vision will require advances in multi-agent coordination, robust interfaces for human–LLM interaction, and domain-specific fine-tuning of models. Concrete research opportunities include designing adaptive orchestration policies that respond to different classes of software, developing hybrid workflows where LLMs collaborate with specialized fuzzing agents, and exploring explainable interfaces that allow humans to oversee high-level strategies without micromanaging execution.  

\textbf{Summary.} Stage~\RomanNum{3} highlights the transformative potential of LLMs for human-in-the-loop fuzzing. 
While Stages~\RomanNum{1} and \RomanNum{2} emphasize improving the scope of monitoring and enabling real-time steering, Stage~\RomanNum{3} envisions an evolution toward synergistic collaboration between humans and intelligent systems. By automating repetitive oversight tasks such as seed deduplication, parameter tuning, and crash triaging, LLMs can substantially reduce the manual burden on testers and accelerate iteration. At the same time, their ability to reason about program semantics, generate structured inputs, and orchestrate adaptive fuzzing strategies positions them to assume responsibilities traditionally reserved for human experts. 
In this trajectory, humans shift into roles of supervision and validation, providing high-level goals and sanity checks, while LLMs execute detailed strategies and continuously adapt workflows. Achieving this vision requires advances in transparency, reliability, and multi-agent coordination, ensuring that LLMs not only reduce cost but also enhance trust, interpretability, and domain-specific adaptability in fuzzing.

\subsection{Concluding Remarks on the Road Ahead.}  
Together, the three stages outlined above sketch a progressive roadmap for advancing human-in-the-loop fuzzing. 
Stage~\RomanNum{1} emphasizes richer and more explainable monitoring, calling for unified observability models and interactive visual analytics that move beyond coarse, outcome-oriented metrics. 
Stage~\RomanNum{2} extends these advances into runtime, integrating human expertise directly into the execution loop through pipelines for in-situ intervention and advanced interaction techniques that make steering both more usable and more effective. 
Stage~\RomanNum{3} looks further ahead, envisioning synergy between humans and LLMs, where intelligent models automate low-level tasks, orchestrate high-level strategies, and emulate human expertise at scale, enabling fuzzing workflows that are adaptive, domain-aware, and intelligence-driven.  

This staged perspective highlights a gradual evolution: from post-execution analysis, to interactive runtime steering, and finally to semi-autonomous collaboration with intelligent systems. By systematically strengthening monitoring, steering, and orchestration, future research can bridge the gap between automated fuzzing and human expertise, paving the way for fuzz testing that is not only more efficient and adaptive, but also more explainable, trustworthy, and aligned with real-world software testing needs.

\section{Conclusion}

This paper presents a systematic review and research roadmap for integrating HITL into fuzz testing. We categorize existing HITL approaches into human monitoring and human steering across the fuzzing lifecycle, revealing significant potential for enhancing interpretability and guiding exploration through expert insight. Looking ahead, we outline a progressive research agenda that evolves from enriched monitoring and interactive steering towards synergistic human-LLM collaboration. This paradigm shift promises to transform fuzzing into a more adaptive, explainable, and powerful vulnerability discovery process, ultimately bridging the gap between automated heuristics and human expertise to secure the next generation of software systems.

\bibliographystyle{ACM-Reference-Format}
\bibliography{TOSEM}

\end{document}